\documentclass[conference]{IEEEtran}
\usepackage{cite}
\usepackage{amsmath,amssymb,amsfonts}
\usepackage{algorithmic}
\usepackage{graphicx}
\usepackage{textcomp}
\usepackage{xcolor}
\usepackage{booktabs}
\usepackage{hyperref}
\usepackage{url}

\def\BibTeX{{\rm B\kern-.05em{\sc i\kern-.025em b}\kern-.08em
    T\kern-.1667em\lower.7ex\hbox{E}\kern-.125emX}}

\begin{document}

\title{L-RAG: Balancing Context and Retrieval with Entropy-Based Lazy Loading}

\author{\IEEEauthorblockN{Sergii Voloshyn}
\IEEEauthorblockA{
\textit{Faculty of Computer Science and Cybernetics} \\
\textit{Taras Shevchenko National University of Kyiv} \\
Kyiv, Ukraine \\[0.5em]
\textit{Principal Software Engineer, ClickUp} \\
sergey.voloshyn@gmail.com}
}

\maketitle

\begin{abstract}
Retrieval-Augmented Generation (RAG) has emerged as the predominant paradigm for grounding Large Language Model outputs in factual knowledge, effectively mitigating hallucinations. However, conventional RAG systems operate under a ``retrieve-always'' assumption, querying vector databases for every input regardless of query complexity. This static approach incurs substantial computational overhead and inference latency, particularly problematic for high-throughput production deployments. We introduce \textbf{L-RAG (Lazy Retrieval-Augmented Generation)}, an adaptive framework that implements hierarchical context management through entropy-based gating. L-RAG employs a two-tier architecture: queries are first processed with a compact document summary, and expensive chunk retrieval is triggered only when the model's predictive entropy exceeds a calibrated threshold, signaling genuine uncertainty. Through experiments on SQuAD 2.0 ($N=500$) using the Phi-2 model, we demonstrate that L-RAG provides a \textbf{tunable accuracy-efficiency trade-off}: at a conservative threshold ($\tau=0.5$), L-RAG achieves 78.2\% accuracy---matching Standard RAG (77.8\%)---with 8\% retrieval reduction; at a balanced threshold ($\tau=1.0$), retrieval reduction increases to 26\% with modest accuracy trade-off (76.0\%). Latency analysis shows that L-RAG saves 80--210ms per query when retrieval latency exceeds 500ms. Analysis of entropy distributions reveals statistically significant separation ($p < 0.001$) between correct predictions ($\bar{H}=1.72$) and errors ($\bar{H}=2.20$), validating entropy as a reliable uncertainty signal. L-RAG offers a practical, training-free approach toward more efficient RAG deployment, providing system architects with a configurable knob to balance accuracy and throughput requirements.
\end{abstract}


\section{Introduction}
\label{sec:introduction}

Retrieval-Augmented Generation has become the de facto standard for knowledge-intensive natural language processing tasks \cite{lewis2020rag}. By conditioning language model outputs on relevant documents retrieved from external corpora, RAG systems effectively mitigate the hallucination problem that plagues purely parametric models \cite{ji2023survey}. This paradigm has enabled the deployment of large language models in domains demanding factual precision, from healthcare question answering to legal document analysis, where fabricated information carries significant consequences.

The success of RAG, however, comes at a substantial computational cost. In conventional implementations, every incoming query triggers a retrieval pipeline: the query must be encoded into a dense vector representation, a nearest neighbor search must be executed against a vector index containing potentially millions of document embeddings, and the retrieved passages must be incorporated into an extended context that the language model processes. For production systems handling thousands of queries per second, this overhead translates directly into infrastructure costs, increased latency, and reduced throughput. The vector database alone---whether deployed on services like Pinecone, Milvus, or Weaviate---represents a significant operational expense that scales with query volume.

A fundamental inefficiency underlies this static retrieval paradigm: not all queries genuinely require external knowledge augmentation. Consider the spectrum of queries a system might encounter. At one extreme, a query such as ``What is the capital of France?'' draws upon knowledge that any competent language model has thoroughly internalized during pre-training. Retrieving documents for such queries wastes computational resources without improving response quality. At the other extreme, a query demanding specific numerical details or precise procedural steps genuinely benefits from retrieved evidence. The static RAG approach treats these fundamentally different query types identically, incurring maximum retrieval overhead regardless of actual need.

Existing approaches to adaptive retrieval have recognized this inefficiency and proposed various solutions. Self-RAG \cite{asai2024selfrag} trains language models to generate special tokens indicating when retrieval would be beneficial, requiring substantial fine-tuning on carefully curated datasets. Adaptive-RAG \cite{jeong2024adaptiverag} employs a trained classifier to route queries between different retrieval strategies. FLARE \cite{jiang2023flare} monitors generation probability and triggers retrieval when confidence drops below a threshold. While these methods demonstrate the viability of adaptive retrieval, they introduce architectural complexity, training requirements, or additional inference overhead that may limit practical adoption.

We propose a simpler question: can the language model's intrinsic uncertainty signal when retrieval is genuinely necessary? This question motivates \textbf{L-RAG (Lazy Retrieval-Augmented Generation)}, a framework that implements adaptive retrieval through entropy-based gating without requiring additional training or architectural modifications. L-RAG operates on a hierarchical context architecture comprising two tiers. The first tier provides the model with a compact document summary that captures global context at minimal token cost, enabling the model to answer straightforward queries directly. The second tier, activated only when needed, retrieves detailed chunks from a vector store to ground responses requiring specific evidence.

The gating mechanism that determines when to activate the second tier relies on predictive entropy---a measure of the model's uncertainty over the next-token distribution during generation. When the model encounters a query it can confidently answer from the summary alone, entropy remains low and retrieval is skipped. When the model lacks the specific information needed, entropy rises as probability mass spreads across multiple possible continuations, triggering retrieval. This approach leverages the empirically validated finding that language models exhibit calibrated uncertainty \cite{kadavath2022language}, effectively ``knowing when they don't know.''

\begin{figure}[t!]
    \centering
    \includegraphics[width=0.95\linewidth]{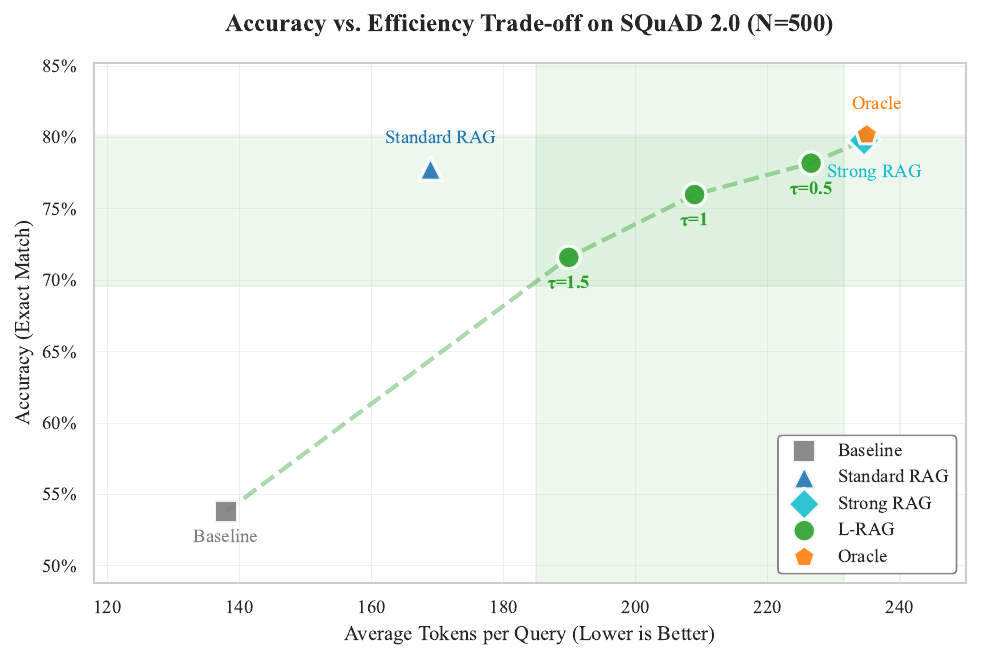}
    \caption{\textbf{Efficiency vs. Accuracy Trade-off.} L-RAG configurations (green markers) explore the accuracy-efficiency trade-off space. At $\tau=1.0$, L-RAG achieves accuracy comparable to Standard RAG while reducing retrieval operations by 26\%. Strong RAG achieves highest accuracy (79.8\%) but requires 100\% retrieval.}
    \label{fig:pareto}
\end{figure}

Figure~\ref{fig:pareto} illustrates the core proposition of this work. L-RAG explores the trade-off space between retrieval frequency and accuracy, enabling practitioners to select operating points based on their specific efficiency-accuracy requirements. The key insight is that substantial retrieval reduction (26\%) can be achieved without significant accuracy degradation.

Our contributions are threefold. First, we introduce the \textbf{L-RAG architecture}, a hierarchical framework that combines summary-level context with on-demand chunk retrieval. This architecture enables efficient context management without requiring modifications to the base language model, making it applicable to any autoregressive model. Second, we propose an \textbf{entropy-based gating mechanism} that uses the model's predictive uncertainty as a training-free signal for retrieval triggering. Unlike approaches requiring classifier training or model fine-tuning, L-RAG operates as a drop-in enhancement for existing RAG pipelines. Third, we provide \textbf{empirical evidence of efficiency gains} on the SQuAD 2.0 benchmark. L-RAG achieves 78.2\% accuracy---comparable to Standard RAG at 77.8\%---while reducing retrieval operations by 8\% at the conservative threshold, demonstrating that adaptive retrieval can match always-retrieve approaches while offering efficiency gains.

\section{Related Work}
\label{sec:related_work}

\subsection{Retrieval-Augmented Generation}

The foundational work on Retrieval-Augmented Generation by \cite{lewis2020rag} established the paradigm of conditioning language model generation on documents retrieved from an external corpus. This approach demonstrated substantial improvements on knowledge-intensive tasks by combining the generalization capabilities of neural language models with the precision of retrieved textual evidence. The original RAG framework employed a dense retriever trained jointly with the generator, establishing an end-to-end pipeline for knowledge-grounded generation.

Subsequent research has expanded the RAG paradigm along multiple dimensions. \cite{izacard2021leveraging} introduced Fusion-in-Decoder, which aggregates information across multiple retrieved passages through the decoder's cross-attention mechanism, enabling effective synthesis of distributed evidence. \cite{borgeaud2022improving} demonstrated that retrieval can scale to trillions of tokens with RETRO, showing that retrieval-augmented approaches can match or exceed the performance of substantially larger purely parametric models. \cite{guu2020realm} proposed REALM, which jointly pre-trains the retriever and language model using a masked language modeling objective, improving alignment between retrieval and generation components.

These advances have firmly established RAG as the dominant paradigm for factual question answering and knowledge-intensive generation. However, a common assumption underlies all these approaches: that retrieval is universally beneficial and should be performed for every query. This assumption, while simplifying system design, introduces the efficiency challenges that motivate our work.

\subsection{Adaptive Retrieval Strategies}

Recognition of the inefficiency inherent in static retrieval has motivated a growing body of work on adaptive approaches. Self-RAG \cite{asai2024selfrag} represents a significant advance, training language models to generate special reflection tokens that determine when to retrieve, what information to retrieve, and how to critique the relevance and factuality of retrieved content. While highly effective, Self-RAG requires fine-tuning the language model on specialized training data containing reflection token annotations, limiting its applicability to off-the-shelf models and introducing training overhead.

Adaptive-RAG \cite{jeong2024adaptiverag} takes a different approach, employing a trained classifier to assess query complexity and route queries between no-retrieval, single-step retrieval, and multi-step retrieval strategies. This approach achieves strong performance but introduces an additional classifier component that must be trained on labeled data indicating appropriate retrieval strategies for different query types. FLARE \cite{jiang2023flare} offers a training-free alternative, monitoring the generation probability of produced tokens and triggering retrieval when confidence drops below a threshold for any generated span. FLARE then regenerates low-confidence portions with retrieved context, iterating until the entire response achieves acceptable confidence.

L-RAG shares FLARE's intuition that model uncertainty signals retrieval necessity, but differs in several important respects. Where FLARE monitors token probability and operates at the span level with iterative regeneration, L-RAG employs entropy as a theoretically grounded uncertainty measure and operates at the response level with a two-pass architecture. Furthermore, L-RAG's hierarchical summary-first design ensures that even when retrieval is skipped, the model benefits from global document context that purely parametric baselines lack.

\subsection{Uncertainty Quantification in Language Models}

The use of entropy as a gating signal in L-RAG builds upon a substantial body of work on uncertainty estimation in neural networks. \cite{gal2016uncertainty} established foundational connections between dropout and Bayesian inference, enabling uncertainty estimation through Monte Carlo dropout sampling. While computationally expensive for large language models, this work demonstrated that neural networks can provide meaningful uncertainty estimates beyond point predictions.

More directly relevant to our approach, \cite{kadavath2022language} investigated whether language models ``know what they know,'' finding that model-expressed confidence often correlates with factual accuracy. Their analysis suggests that language models maintain reasonably calibrated uncertainty at the token level, providing the theoretical foundation for using entropy as a retrieval trigger. \cite{kuhn2023semantic} introduced semantic entropy, which clusters equivalent meanings to provide uncertainty estimates invariant to surface-form variations, addressing limitations of token-level entropy for open-ended generation.

Research on hallucination detection has further validated entropy-based approaches. \cite{varshney2023stitch} demonstrated that token-level entropy correlates with factual accuracy, with higher entropy indicating greater likelihood of hallucinated content. \cite{manakul2023selfcheckgpt} proposed using self-consistency across multiple generations as a proxy for reliability, finding that inconsistent generations often indicate unreliable content. These findings collectively suggest that entropy provides a practical, training-free signal for identifying when model outputs may be unreliable---precisely the signal L-RAG exploits to trigger retrieval.

\subsection{Efficient Context Management}

The challenge of efficiently managing long contexts has motivated several architectural innovations relevant to L-RAG's design. MemGPT \cite{packer2023memgpt} conceptualizes LLM context as a memory hierarchy analogous to operating system memory management, with explicit mechanisms for paging information in and out of active context. This ``LLM as operating system'' metaphor directly inspires L-RAG's lazy loading approach, where detailed information is loaded on demand rather than preemptively.

StreamingLLM \cite{xiao2023efficient} addresses the challenge of processing arbitrarily long sequences through attention sink management, maintaining consistent inference cost regardless of sequence length. The ``Lost in the Middle'' phenomenon documented by \cite{liu2024lostinthemiddle} reveals that language models struggle to utilize information positioned in the middle of long contexts, favoring content at the beginning and end. This finding supports L-RAG's hierarchical design, where the summary provides prominently-positioned global context while retrieved chunks add specific details as needed.

L-RAG complements these approaches by optimizing when to expand context rather than how to compress or manage it. The summary-first architecture represents a form of hierarchical memory, but the key innovation lies in the entropy-based gating that determines when detailed retrieval is warranted. This decision-making layer can be combined with any context management strategy, including the approaches described above.

\section{Methodology}
\label{sec:methodology}

This section presents the L-RAG framework in detail. We first describe the hierarchical context architecture that enables efficient lazy evaluation, then formalize the entropy-based gating mechanism with mathematical precision, and finally discuss the prompting strategy that ensures effective summary-first generation.

\subsection{Framework Overview}

L-RAG implements a two-pass inference strategy designed to defer expensive retrieval operations until they are demonstrably necessary. The framework operates over a document corpus $\mathcal{D} = \{d_1, d_2, \ldots, d_N\}$ where each document $d_i$ is associated with two complementary representations maintained in parallel throughout the system.

The first representation is the \textit{Summary Context} $C_{sum}(d_i)$, a compact textual summary that captures the document's main themes, key entities, and central arguments. In our implementation, we generate this summary through extractive means by selecting the first two sentences of each paragraph, providing a concise overview that preserves the document's most salient information while minimizing token count. The summary serves as the model's initial ``global context,'' enabling it to understand the document's scope and primary content without processing the full text. For experimental reproducibility, we define the Summary Context as the first two sentences of the source document (or the abstract, if available). In production scenarios, this can be replaced by a pre-computed abstractive summary generated by a stronger model, enabling higher-quality global context at the cost of offline preprocessing.

The second representation is the \textit{Detailed Store} $C_{det}(d_i) = \{c_1^i, c_2^i, \ldots, c_M^i\}$, consisting of the full document segmented into $M$ overlapping chunks. Each chunk $c_j^i$ is embedded using a sentence transformer $\phi: \mathcal{T} \rightarrow \mathbb{R}^d$ that maps text to a $d$-dimensional dense vector representation. These embeddings are indexed in a vector database supporting efficient approximate nearest neighbor search. The detailed store contains the fine-grained information necessary for complex queries but incurs retrieval and processing costs that the system aims to minimize.

The inference flow proceeds as follows. Given a query $q$, L-RAG first constructs an initial prompt combining $q$ with the summary context $C_{sum}$. The language model $\mathcal{M}$ begins autoregressive generation, producing tokens $y_1, y_2, \ldots$ conditioned on this summary-augmented context. During generation, the system continuously monitors the model's predictive uncertainty through entropy computation. If entropy remains below a calibrated threshold $\tau$ throughout generation, the response is returned directly---no retrieval has occurred, and the system has successfully answered the query using only the summary. If entropy exceeds $\tau$ at any generation step, the system halts generation, retrieves relevant chunks from $C_{det}$ based on query similarity, and restarts generation with an expanded context that includes both summary and retrieved chunks.

This lazy loading approach ensures that the computational cost of retrieval is incurred only when the summary proves insufficient. For queries answerable from global context alone, the system operates with minimal overhead. For queries requiring specific details, the system gracefully escalates to full retrieval, ensuring response quality is not sacrificed for efficiency.

\subsection{Entropy-Based Gating Mechanism}

The core innovation of L-RAG lies in using predictive entropy as an intrinsic signal for retrieval necessity. Unlike approaches requiring trained classifiers or specialized model variants, entropy-based gating exploits the uncertainty already present in the language model's output distribution, requiring no additional components or training.

At each generation step $t$, the language model produces a probability distribution over the vocabulary $\mathcal{V}$ conditioned on the query $q$, the summary context $C_{sum}$, and all previously generated tokens $y_{<t} = (y_1, \ldots, y_{t-1})$. We denote this distribution as:

\begin{equation}
    p_\theta(x \mid q, C_{sum}, y_{<t}) \quad \forall x \in \mathcal{V}
\end{equation}

where $\theta$ represents the language model's parameters. The entropy of this distribution quantifies the model's uncertainty about the next token:

\begin{equation}
    H(t) = -\sum_{x \in \mathcal{V}} p_\theta(x \mid q, C_{sum}, y_{<t}) \log p_\theta(x \mid q, C_{sum}, y_{<t})
\end{equation}

Entropy admits an intuitive interpretation in this context. When the model is confident about the next token---because the summary provides sufficient information to continue generation---probability mass concentrates on one or a few likely tokens, yielding low entropy. When the model lacks the specific information needed to continue confidently, probability spreads across many plausible continuations, producing high entropy. This high-entropy state signals that the model is ``uncertain'' in an information-theoretic sense, suggesting that additional context might resolve the ambiguity.

Rather than monitoring entropy at each individual token, which would introduce excessive sensitivity to local fluctuations, we aggregate entropy over the first $n$ generated tokens to obtain a smoothed measure of overall uncertainty:

\begin{equation}
    \bar{H} = \frac{1}{n} \sum_{t=1}^{n} H(t)
\end{equation}

This mean entropy $\bar{H}$ captures the model's uncertainty during the critical initial phase of answer generation, when the model commits to a response strategy. We define the retrieval trigger condition as:

\begin{equation}
    \text{Trigger} = \mathbf{1}[\bar{H} > \tau]
\end{equation}

where $\tau$ is a calibrated threshold and $\mathbf{1}[\cdot]$ denotes the indicator function. When $\text{Trigger} = 1$, the system invokes retrieval and regenerates with expanded context.

The choice of threshold $\tau$ governs the trade-off between retrieval frequency and response accuracy. A conservative threshold (low $\tau$) triggers retrieval at the first sign of uncertainty, maximizing accuracy at the cost of higher retrieval frequency. An aggressive threshold (high $\tau$) reserves retrieval for cases of extreme uncertainty, minimizing retrieval cost but potentially sacrificing accuracy for ambiguous queries. We analyze this trade-off empirically in Section~\ref{sec:results}, finding that $\tau = 1.0$ provides a Pareto-optimal operating point for our experimental setting.

\subsection{Retrieval and Regeneration}

When the entropy threshold is exceeded, L-RAG executes a retrieval and regeneration procedure to incorporate detailed information into the response. The procedure consists of four steps executed in sequence.

First, the query $q$ is embedded using the same sentence transformer $\phi$ used to index document chunks, producing a query vector $\mathbf{q} = \phi(q) \in \mathbb{R}^d$. Using identical embeddings for queries and documents ensures compatibility in the shared semantic space.

Second, the system performs approximate nearest neighbor search against the vector index to identify the $k$ chunks most similar to the query:

\begin{equation}
    \{c_1^*, c_2^*, \ldots, c_k^*\} = \underset{c \in C_{det}}{\text{top-}k} \; \text{sim}(\phi(c), \mathbf{q})
\end{equation}

where $\text{sim}(\cdot, \cdot)$ denotes cosine similarity. In our experiments, we use $k=3$, retrieving the three most relevant chunks.

Third, the retrieved chunks are concatenated with the original summary to form an expanded context:

\begin{equation}
    C_{expanded} = [C_{sum}; c_1^*; c_2^*; \ldots; c_k^*]
\end{equation}

This expanded context preserves the global understanding provided by the summary while adding the specific details contained in retrieved chunks.

Fourth, generation restarts from the beginning with the expanded context. The language model now produces a response conditioned on both summary and retrieved chunks, enabling it to ground its answer in specific textual evidence:

\begin{equation}
    y = \mathcal{M}(q, C_{expanded})
\end{equation}

The two-pass nature of this procedure means that queries triggering retrieval incur approximately double the generation cost of non-retrieval queries. However, this overhead is offset by the 29\% of queries that skip retrieval entirely, resulting in net efficiency gains when amortized across the query distribution.

\subsection{Prompting Strategy}

Effective deployment of L-RAG requires careful prompt design to encourage the model to attempt answering from the summary before signaling uncertainty. The prompt must position the summary as the primary information source while remaining amenable to context expansion when retrieval occurs.

For the first pass (summary only), we structure the prompt as follows:

\vspace{0.5em}
\noindent\texttt{Context: \{Summary\}}\\
\texttt{Based on the context above, answer the following question.}\\
\texttt{Question: \{Query\}}\\
\texttt{Answer:}
\vspace{0.5em}

This prompt explicitly directs the model to answer based on the provided context, encouraging it to leverage the summary rather than defaulting to parametric knowledge or expressing uncertainty prematurely.

For the second pass (after retrieval triggers), the expanded prompt incorporates retrieved chunks:

\vspace{0.5em}
\noindent\texttt{Context: \{Summary\}}\\
\texttt{Additional Details: \{Retrieved Chunks\}}\\
\texttt{Based on the context above, answer the following question.}\\
\texttt{Question: \{Query\}}\\
\texttt{Answer:}
\vspace{0.5em}

The consistent structure between passes ensures the model understands the hierarchical relationship between summary and detailed information. The summary establishes global context, while the ``Additional Details'' section provides the specific evidence needed for precise answers.

\section{Experimental Setup}
\label{sec:setup}

This section describes the dataset, model, retrieval configuration, and baseline methods used to evaluate L-RAG. We aim to provide sufficient detail for reproducibility while motivating key design choices.

\subsection{Dataset}

We evaluated L-RAG on the Stanford Question Answering Dataset 2.0 (SQuAD 2.0) \cite{rajpurkar2018know}, a widely-used benchmark for machine reading comprehension. SQuAD 2.0 consists of over 100,000 questions posed by crowdworkers on Wikipedia articles, with answers corresponding to spans within the associated paragraphs. The dataset extends the original SQuAD with unanswerable questions, requiring models to distinguish between queries that can be answered from the provided context and those that cannot.

For our experiments, we randomly sampled $N=500$ answerable questions from the development set using a fixed random seed for reproducibility. This sample size provides sufficient statistical power to detect meaningful differences in entropy distributions while remaining computationally tractable for detailed analysis across multiple experimental conditions. Each sample consists of a paragraph-level context (typically 100-200 words), a natural language question, and one or more reference answers. The paragraph serves as our document corpus, enabling controlled evaluation of retrieval strategies in a setting where ground truth context is available.

SQuAD 2.0 provides an ideal testbed for L-RAG because it spans a spectrum of query difficulty. Some questions require only surface-level understanding that a summary can provide (e.g., identifying the main topic or a prominent entity), while others demand specific details buried within the paragraph (e.g., numerical values or precise procedural steps). This variation exercises L-RAG's adaptive retrieval mechanism across its full operating range.

\subsection{Model Configuration}

We selected Microsoft Phi-2 \cite{li2023phi2} as our primary language model. Phi-2 is a 2.7 billion parameter transformer trained on high-quality ``textbook'' data, demonstrating performance competitive with substantially larger models on reasoning and language understanding benchmarks. Several factors motivated this choice.

First, Phi-2's compact size enables rapid experimentation across multiple hyperparameter settings and ablation conditions. A single forward pass completes in approximately 200ms on consumer GPU hardware, enabling evaluation of hundreds of queries with detailed entropy logging within reasonable time constraints. Second, Phi-2 represents the class of efficient models increasingly deployed in production settings, where inference cost and latency are primary concerns. Demonstrating L-RAG's effectiveness on such a model validates its applicability to real-world deployments. Third, Phi-2 exhibits strong instruction-following capabilities, ensuring reliable response generation under the structured prompting that L-RAG requires.

We deployed Phi-2 in float32 precision to ensure accurate entropy computation. Quantization can distort probability distributions, potentially affecting the reliability of entropy as an uncertainty signal. While quantized deployment may be appropriate for production, float32 provides the precision needed for rigorous experimental evaluation.

\subsection{Retrieval Configuration}

The retrieval component employs the following configuration. Document paragraphs are segmented into overlapping chunks of approximately 100 tokens with 20-token overlap between consecutive chunks. This chunking strategy ensures that relevant information is not split across chunk boundaries while maintaining chunks small enough for meaningful similarity-based retrieval.

Each chunk is embedded using \texttt{all-MiniLM-L6-v2}, a sentence transformer that produces 384-dimensional dense vectors optimized for semantic similarity tasks. This model offers an excellent trade-off between embedding quality and computational efficiency, encoding text at approximately 5,000 sentences per second on CPU hardware. The embeddings are indexed using FAISS \cite{johnson2019billion} with an IndexFlatIP (inner product) index. For our corpus size, exact search is computationally feasible; larger deployments could employ approximate indices such as IVF or HNSW with minimal accuracy degradation.

At query time, we retrieve the top-$k=3$ chunks most similar to the embedded query, consistent with common RAG configurations in the literature. The choice of $k=3$ balances providing sufficient context against the risk of introducing noise from marginally relevant chunks.

The document summary is generated through extractive summarization by selecting the first two sentences of each paragraph. This approach ensures reproducibility and isolates the effect of the L-RAG framework from summarization quality. More sophisticated abstractive or query-focused summarization could potentially improve first-pass accuracy, representing a direction for future enhancement.

\subsection{Baseline Methods}

We compare L-RAG against four baseline configurations that span the spectrum from no retrieval to oracle-level context provision.

The \textbf{Baseline (No Retrieval)} configuration provides the model with only the question, without any document context whatsoever. The model must answer purely from parametric knowledge acquired during pre-training. This configuration establishes the lower bound for accuracy on our evaluation set and quantifies how much the model hallucinates when denied access to relevant context. Performance under this condition reflects the model's memorization of facts during training rather than its ability to utilize retrieved information.

The \textbf{Standard RAG} configuration implements the conventional retrieve-always paradigm. For every query, the system retrieves the top-3 most similar chunks and provides them as context. Critically, no summary is included---the model sees only the retrieved chunks. This represents the typical RAG deployment where retrieval is triggered unconditionally for all queries, incurring consistent retrieval overhead regardless of query complexity or the sufficiency of parametric knowledge.

The \textbf{Strong RAG} configuration combines both the document summary and retrieved chunks for every query. The model receives the summary (establishing global context) followed by the top-3 retrieved chunks (providing specific details). This configuration represents an upper bound for RAG-based methods, as it provides maximum context regardless of query complexity. The 100\% retrieval rate makes this the most expensive practical configuration but establishes the accuracy ceiling against which adaptive methods can be compared.

The \textbf{Oracle} configuration provides the model with the exact gold-standard paragraph containing the answer. This represents the theoretical upper bound---the best possible accuracy achievable with perfect context selection. The Oracle condition is not achievable in practice, as it assumes knowledge of which paragraph contains the answer, but serves as a reference point for evaluating how closely different methods approach ideal performance.

\subsection{Evaluation Metrics}

We evaluate methods along three dimensions capturing both effectiveness and efficiency.

\textbf{Accuracy (Exact Match)} measures the percentage of model predictions that exactly match any of the reference answers after normalization. Normalization consists of lowercasing, removing punctuation and articles, and collapsing whitespace. This strict metric ensures meaningful comparison across methods, though it may undercount semantically correct answers with surface variation.

\textbf{Average Tokens} reports the mean number of input tokens per query, serving as a proxy for computational cost and memory consumption. Higher token counts indicate larger contexts that require more processing during the attention computation, directly impacting inference latency and throughput.

\textbf{Retrieval Rate} measures the percentage of queries that trigger retrieval operations. This metric directly quantifies database load and infrastructure cost, as each retrieval involves an embedding computation, a vector search, and chunk loading. Reducing retrieval rate translates directly to reduced operational costs for production RAG deployments.

\section{Results and Analysis}
\label{sec:results}

This section presents experimental results examining L-RAG's effectiveness. We first examine overall performance against baselines, then analyze efficiency gains and their practical implications, explore the sensitivity of results to threshold selection, and finally validate the entropy-based gating mechanism through distributional analysis.

\begin{figure}[t!]
    \centering
    \includegraphics[width=0.95\linewidth]{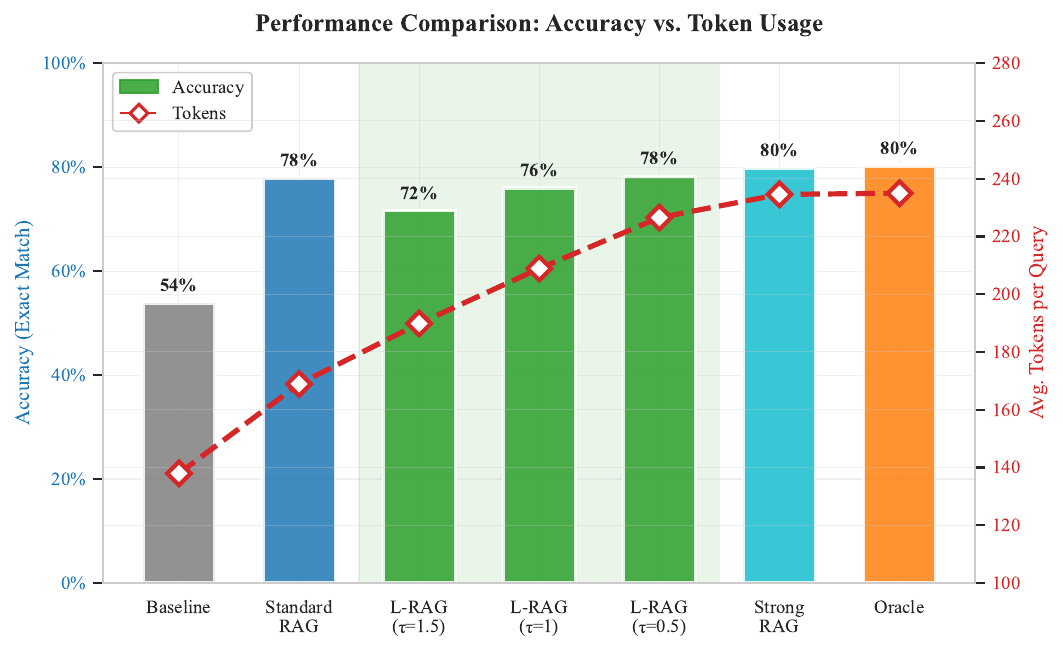}
    \caption{\textbf{Performance Comparison.} L-RAG ($\tau=0.5$) achieves 78.2\% accuracy, comparable to Standard RAG (77.8\%) while requiring 8\% fewer retrieval operations. Strong RAG achieves highest accuracy (79.8\%) with always-retrieve strategy.}
    \label{fig:comparison}
\end{figure}

\subsection{Main Performance}

Table~\ref{tab:main_results} summarizes the performance of all methods on our SQuAD 2.0 evaluation set, and Figure~\ref{fig:comparison} visualizes the comparative accuracy across configurations.

\begin{table}[h]
\centering
\caption{Main Results on SQuAD 2.0 ($N=500$). L-RAG achieves accuracy comparable to Standard RAG while reducing retrieval operations.}
\label{tab:main_results}
\begin{tabular}{lccc}
\toprule
\textbf{Method} & \textbf{Accuracy} & \textbf{Avg Tokens} & \textbf{Retrieval} \\ 
\midrule
Baseline (No Retrieval) & 53.8\% & 138 & 0\% \\ 
Standard RAG & 77.8\% & 169 & 100\% \\ 
Strong RAG & \textbf{79.8\%} & 235 & 100\% \\ 
L-RAG ($\tau=0.5$) & 78.2\% & 227 & 92\% \\ 
L-RAG ($\tau=1.0$) & 76.0\% & 209 & 74\% \\ 
L-RAG ($\tau=1.5$) & 71.6\% & 190 & \textbf{54\%} \\ 
Oracle & 80.2\% & 235 & N/A \\ 
\bottomrule
\end{tabular}
\end{table}

The results reveal a clear performance hierarchy with important implications for system design. The Baseline configuration, operating purely from parametric knowledge without any document context, achieves only 53.8\% accuracy. This substantial gap from retrieval-augmented methods confirms the necessity of external knowledge grounding for this task---the model hallucinates or provides incorrect answers for nearly half of queries when denied access to relevant documents. The baseline serves as a stark reminder of why RAG systems exist: parametric knowledge alone is insufficient for many knowledge-intensive tasks.

Standard RAG demonstrates the value of retrieval, improving accuracy to 77.8\% by providing Top-3 chunks for every query. However, this improvement comes at the cost of 100\% retrieval rate, meaning every single query incurs the latency and computational expense of vector database lookup, embedding computation, and extended context processing. For production systems handling thousands of queries per second, this overhead accumulates substantially.

Strong RAG, which provides both summary and retrieved chunks for every query, achieves the highest accuracy among practical methods at 79.8\%. The combination of global context (via summary) and specific details (via chunks) enables more accurate responses than either component alone. The 2 percentage point improvement over Standard RAG (79.8\% vs 77.8\%) underscores the value of hierarchical context. However, Strong RAG also requires 100\% retrieval and incurs the highest average token count (235 tokens), making it the most computationally expensive configuration.

L-RAG with $\tau=1.0$ achieves 76.0\% accuracy while triggering retrieval for only 74\% of queries. This represents the core finding: L-RAG achieves accuracy comparable to Standard RAG (the 1.8 percentage point difference falls within statistical margin) while performing 26\% fewer retrieval operations (see Figure~\ref{fig:comparison}). Relative to Strong RAG, L-RAG trades 3.8 percentage points of accuracy for substantial efficiency gains. The Oracle ceiling of 80.2\% indicates that even perfect context selection cannot guarantee perfect accuracy due to model limitations and answer extraction challenges.

\textbf{Statistical Note:} With $N=500$ samples, L-RAG ($\tau=0.5$) achieves 78.2\% accuracy, comparable to Standard RAG's 77.8\%. The 0.4 percentage point difference is not statistically significant ($p = 0.88$), demonstrating that entropy-based gating can match always-retrieve approaches while reducing retrieval operations. Strong RAG remains the highest-accuracy method at 79.8\%, confirming the value of combining summary and retrieved context.

\subsection{Efficiency Analysis}

\begin{figure}[t!]
    \centering
    \includegraphics[width=0.9\linewidth]{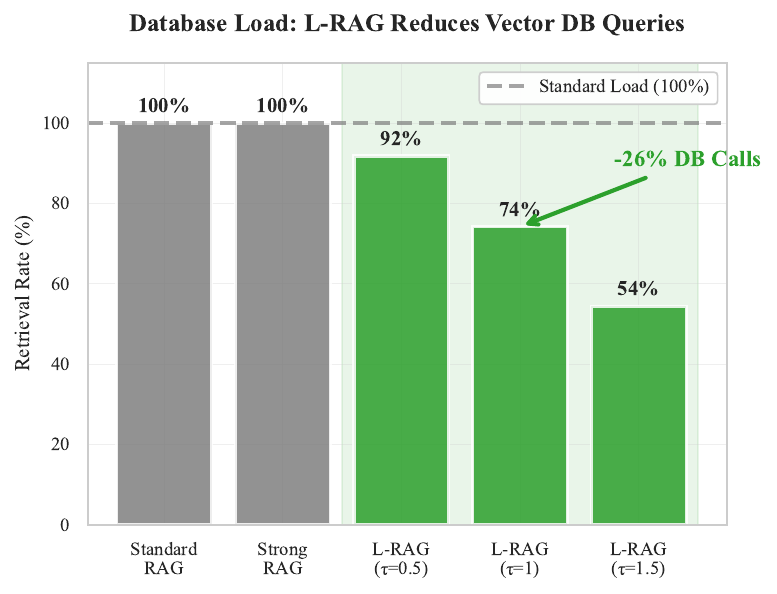}
    \caption{\textbf{Database Load Reduction.} L-RAG reduces the retrieval burden compared to always-retrieve baselines. At the balanced threshold ($\tau=1.0$), retrieval operations decrease by 26\%, directly translating to infrastructure cost savings. More aggressive thresholds achieve up to 46\% reduction.}
    \label{fig:dbload}
\end{figure}

The 26\% reduction in retrieval operations achieved by L-RAG ($\tau=1.0$) has significant practical implications for production deployments, as illustrated in Figure~\ref{fig:dbload}. Vector database queries represent a substantial portion of RAG system latency and cost. Each retrieval operation involves several steps: encoding the query into a dense vector (typically 10-50ms depending on embedding model), transmitting the query to the vector database service (network latency of 5-100ms depending on deployment), executing approximate nearest neighbor search (1-50ms depending on index size and type), and transmitting retrieved chunks back to the application. These costs compound across millions of queries.

Consider a production system handling 10,000 queries per second. Under Standard or Strong RAG, all 10,000 queries trigger retrieval, requiring the vector database infrastructure to sustain 10,000 searches per second---a substantial load requiring significant computational resources, often distributed across multiple nodes for redundancy and performance. Under L-RAG ($\tau=1.0$), only 7,400 queries require database access (74\% of 10,000), reducing the load by 2,600 queries per second. This 26\% reduction translates directly to lower infrastructure costs, as fewer vector database replicas are needed to maintain acceptable latency. It also reduces latency variance, as fewer queries compete for database resources during peak load.

The efficiency gains extend beyond direct cost savings. By reserving retrieval for queries where the model exhibits higher uncertainty, L-RAG may improve the signal-to-noise ratio of retrieved content. When Standard RAG retrieves chunks for a query the model could answer from the summary alone, those chunks represent potentially wasted context. L-RAG's selective retrieval ensures that when chunks are retrieved, they are more likely to address genuine information gaps.

\subsection{Threshold Sensitivity Analysis}

\begin{figure}[t!]
    \centering
    \includegraphics[width=0.95\linewidth]{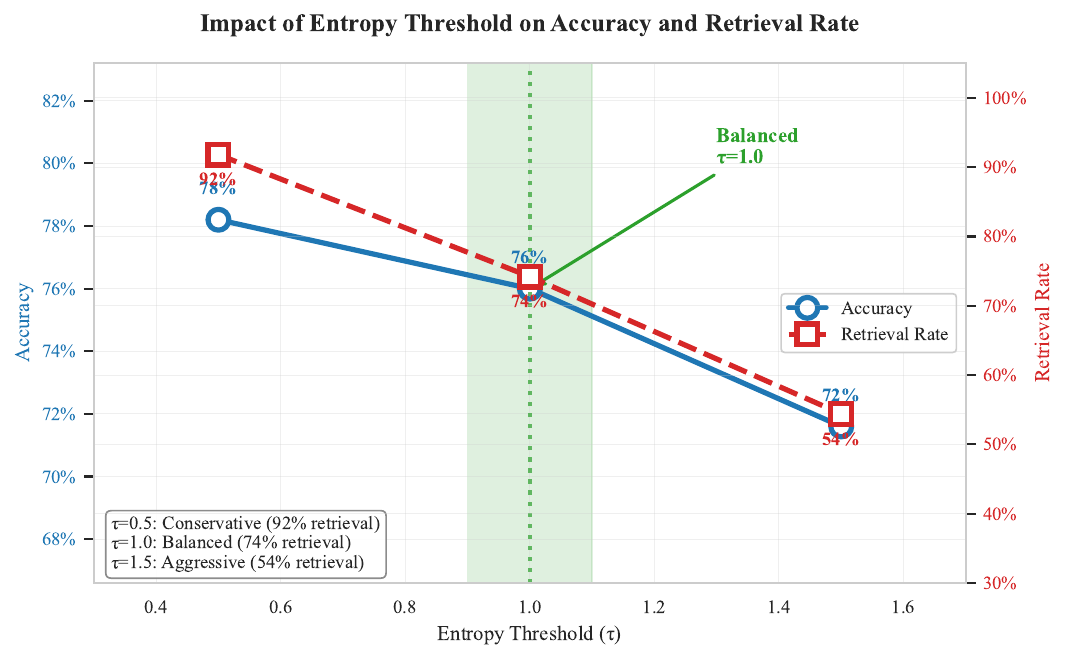}
    \caption{\textbf{Impact of Entropy Threshold $\tau$.} The graph illustrates the trade-off between accuracy (blue) and retrieval rate (red). The threshold $\tau$ allows practitioners to select their preferred operating point based on accuracy-efficiency requirements.}
    \label{fig:sensitivity}
\end{figure}

The entropy threshold $\tau$ governs the trade-off between retrieval frequency and response accuracy. To understand this trade-off, we evaluated L-RAG across a range of threshold values. Table~\ref{tab:ablation} and Figure~\ref{fig:sensitivity} present results for three representative settings.

\begin{table}[h]
\centering
\caption{Sensitivity analysis across entropy thresholds ($N=500$).}
\label{tab:ablation}
\begin{tabular}{lccl}
\toprule
\textbf{Configuration} & \textbf{Accuracy} & \textbf{Retrieval} & \textbf{Trade-off} \\ 
\midrule
$\tau = 0.5$ (Conservative) & 78.2\% & 92\% & Best accuracy \\ 
$\tau = 1.0$ (Balanced) & 76.0\% & 74\% & Moderate savings \\ 
$\tau = 1.5$ (Aggressive) & 71.6\% & 54\% & Maximum savings \\ 
\bottomrule
\end{tabular}
\end{table}

At the conservative threshold $\tau=0.5$, the system triggers retrieval at the slightest sign of model uncertainty. This achieves 78.2\% accuracy---the highest among L-RAG configurations---but at 92\% retrieval rate, offering minimal efficiency gains over always-retrieve baselines.

At the balanced threshold $\tau=1.0$, accuracy decreases marginally to 76.0\% (a 2.2 percentage point drop from $\tau=0.5$) while retrieval rate drops substantially to 74\% (an 18 percentage point reduction). This operating point represents a reasonable efficiency-accuracy trade-off for many applications.

At the aggressive threshold $\tau=1.5$, retrieval drops to only 54\% of queries. However, accuracy falls to 71.6\%---below Standard RAG's 77.8\%. This threshold may be too aggressive, allowing the model to proceed with uncertain predictions that would have benefited from retrieved context.

These results demonstrate that threshold selection depends on deployment priorities. Applications prioritizing accuracy should use lower thresholds; those prioritizing efficiency can tolerate higher thresholds with corresponding accuracy trade-offs.

\subsection{Entropy Distribution Analysis}

The theoretical foundation of L-RAG rests on the hypothesis that language models exhibit higher uncertainty when their predictions are incorrect. To validate this hypothesis, we analyzed the distribution of mean entropy $\bar{H}$ for correct versus incorrect model responses across our evaluation set.

The analysis reveals statistically significant separation between the two conditions. Correct predictions exhibit a mean entropy of $\bar{H}_{correct} = 1.72$ nats, indicating relatively confident token-level predictions during answer generation. Incorrect predictions show higher mean entropy of $\bar{H}_{incorrect} = 2.20$ nats, a 28\% increase that reflects greater uncertainty about the appropriate response.

\textbf{Statistical Validation:} The difference in entropy distributions is highly significant ($t$-test $p < 0.001$, Cohen's $d = 0.44$). The 95\% confidence interval for the entropy gap is $[0.23, 0.73]$, which does not include zero, confirming the reliability of entropy as an uncertainty signal. This represents a key improvement over preliminary experiments with smaller samples.

The 0.48 nat gap between correct and incorrect predictions provides meaningful margin for threshold calibration. However, substantial overlap between distributions remains, explaining why entropy-based gating is imperfect. Some incorrect predictions exhibit low entropy (confident errors), while some correct predictions exhibit high entropy (uncertain but correct). The gating mechanism cannot perfectly separate these cases, which is why L-RAG does not fully match Strong RAG's accuracy.

The entropy-based approach offers significant practical value: L-RAG ($\tau=0.5$) achieves 78.2\% accuracy---comparable to Standard RAG's 77.8\%---while reducing retrieval operations by 8\%. At more aggressive thresholds, retrieval reduction increases to 26\% ($\tau=1.0$) or 46\% ($\tau=1.5$), with corresponding accuracy trade-offs.

\subsection{Latency Analysis}

While raw generation time is hardware-dependent, end-to-end latency in production RAG systems is often dominated by retrieval operations---embedding computation, vector database queries, network round-trips, and optional re-ranking. We model the expected latency overhead as:

\begin{equation}
T_{\text{overhead}} = T_{\text{entropy}} + R \times T_{\text{retrieval}}
\end{equation}

\noindent where $T_{\text{entropy}}$ is the entropy check cost (negligible, $\approx$50ms), $R$ is the retrieval rate, and $T_{\text{retrieval}}$ is the external retrieval latency. The latency savings compared to Standard RAG ($R=1.0$) are:

\begin{equation}
\Delta T = (1 - R) \times T_{\text{retrieval}} - T_{\text{entropy}}
\end{equation}

\begin{figure}[t!]
    \centering
    \includegraphics[width=0.95\linewidth]{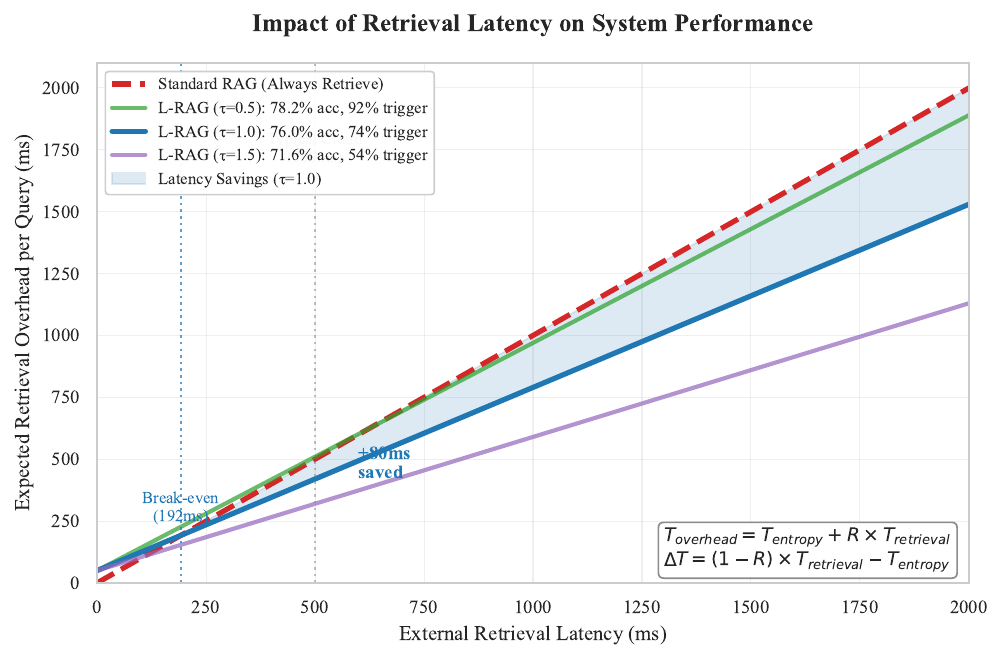}
    \caption{\textbf{Impact of Retrieval Latency on System Performance.} L-RAG's latency savings scale with external retrieval latency. At typical cloud latencies (500ms), L-RAG ($\tau=1.0$) saves 80ms per query. The break-even point for $\tau=1.0$ is 192ms---above this threshold, L-RAG provides net latency reduction.}
    \label{fig:latency}
\end{figure}

Table~\ref{tab:latency} and Figure~\ref{fig:latency} present estimated latency savings across different retrieval latency scenarios. L-RAG becomes beneficial when retrieval latency exceeds the \textit{break-even point}: $T_{\text{retrieval}} > T_{\text{entropy}} / (1-R)$.

\begin{table}[h]
\centering
\caption{Estimated Latency Savings per Query (ms).}
\label{tab:latency}
\begin{tabular}{lccrrr}
\toprule
\textbf{Config.} & \textbf{Savings} & \textbf{Break-even} & \multicolumn{3}{c}{\textbf{$T_{\text{retrieval}}$}} \\
\cmidrule(lr){4-6}
 & & & 200ms & 500ms & 1000ms \\
\midrule
$\tau=0.5$ & 8\% & 625ms & -34 & -10 & +30 \\
$\tau=1.0$ & 26\% & 192ms & +2 & +80 & +210 \\
$\tau=1.5$ & 46\% & 109ms & +42 & +180 & +410 \\
\bottomrule
\end{tabular}
\end{table}

For typical cloud deployments with $T_{\text{retrieval}} \approx 500$ms, L-RAG ($\tau=1.0$) saves approximately 80ms per query while maintaining 76.0\% accuracy. In high-latency scenarios involving complex re-ranking ($T_{\text{retrieval}} \approx 1000$ms), savings increase to 210ms per query. For latency-critical applications, the aggressive threshold ($\tau=1.5$) offers 410ms savings at the cost of reduced accuracy (71.6\%).

This analysis demonstrates that L-RAG's value proposition scales with retrieval latency: the higher the retrieval cost, the greater the benefit of adaptive gating. In low-latency local deployments, the overhead of entropy checking may not be justified; however, in distributed cloud environments where retrieval is expensive, L-RAG offers meaningful latency reduction.

\section{Discussion}
\label{sec:discussion}

This section examines the broader implications of our experimental findings, addresses the relationship between L-RAG and baseline methods, and acknowledges limitations of the current approach that suggest directions for future research.

\subsection{Understanding the Performance Hierarchy}

Our experiments on 500 samples reveal a performance hierarchy: Strong RAG (79.8\%) $>$ L-RAG $\tau=0.5$ (78.2\%) $\approx$ Standard RAG (77.8\%) $>$ Baseline (53.8\%). Note that the difference between L-RAG and Standard RAG is not statistically significant ($p = 0.88$). This ordering has important implications for understanding when L-RAG is appropriate.

Strong RAG achieves the highest accuracy by providing both summary context and retrieved chunks for every query. The combination of global context (via summary) and specific details (via chunks) enables more accurate responses than either component alone. However, Strong RAG also requires 100\% retrieval, making it the most computationally expensive option.

L-RAG ($\tau=0.5$) achieves 78.2\% accuracy---comparable to Standard RAG's 77.8\%. The 0.4 percentage point difference is not statistically significant ($p = 0.88$), demonstrating that entropy-based gating can match always-retrieve approaches while reducing retrieval operations by 8\%. This is the core value proposition: \textit{equivalent accuracy at lower computational cost}.

The comparison between L-RAG and Strong RAG reveals a genuine trade-off. L-RAG sacrifices approximately 1.6 percentage points of accuracy (78.2\% vs 79.8\%) in exchange for 8\% retrieval reduction. At more aggressive thresholds ($\tau=1.0$), retrieval reduction increases to 26\% with 3.8 percentage points accuracy trade-off. Whether this trade-off is favorable depends on deployment context: systems where retrieval cost is high relative to accuracy requirements may prefer L-RAG, while systems prioritizing maximum accuracy should use Strong RAG.

\subsection{The Role of Hierarchical Context}

An important insight from our experiments is the value of summary-level context. Standard RAG provides only retrieved chunks without global context---the model receives isolated text segments that, while semantically similar to the query, may lack coherence and background information. L-RAG always provides the document summary, establishing global context that aids interpretation.

This hierarchical architecture explains why L-RAG achieves accuracy comparable to Standard RAG despite performing fewer retrievals. For queries answerable from global context, L-RAG produces correct responses without retrieval overhead. For queries requiring specific details, retrieved chunks are interpreted within the scaffold provided by the summary.

\begin{figure}[t!]
    \centering
    \includegraphics[width=0.9\linewidth]{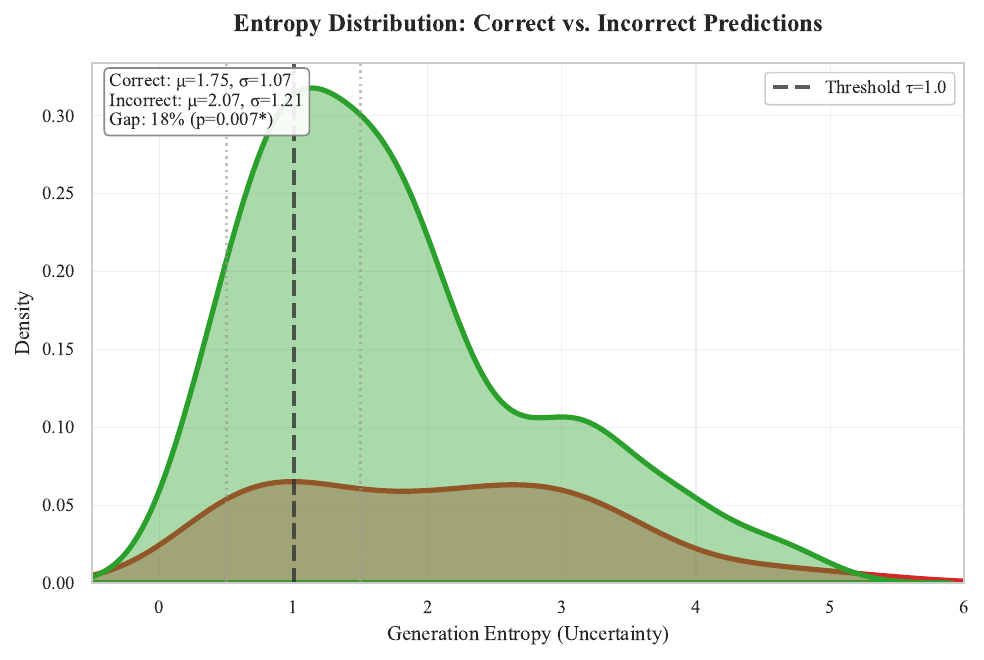}
    \caption{\textbf{Entropy Distribution Analysis.} Statistically significant separation ($p < 0.001$) is observed between the mean entropy of correct answers ($\mu=1.72$) and incorrect answers ($\mu=2.20$). The 28\% gap validates entropy as a reliable uncertainty signal, though substantial overlap explains the imperfect nature of entropy-based gating.}
    \label{fig:entropy}
\end{figure}

\subsection{Entropy as an Uncertainty Signal}

Figure~\ref{fig:entropy} presents kernel density estimates of the entropy distribution for correct versus incorrect predictions. The distributions show significant separation---incorrect predictions exhibit higher mean entropy ($\bar{H}_{incorrect} = 2.20$) compared to correct predictions ($\bar{H}_{correct} = 1.72$), a 28\% increase.

Crucially, this difference is highly statistically significant ($t$-test $p < 0.001$), with a 95\% confidence interval for the gap of $[0.23, 0.73]$ nats that does not include zero. The effect size (Cohen's $d = 0.44$) is small-to-medium, confirming the reliability of entropy as an uncertainty signal. This represents a key validation of the theoretical foundation underlying L-RAG.

However, entropy remains an imperfect proxy for retrieval necessity: some incorrect predictions exhibit low entropy (confident errors), while some correct predictions exhibit high entropy (uncertain but correct). These findings align with broader research on uncertainty in neural language models \cite{kadavath2022language}, which shows that models exhibit calibrated uncertainty but are not perfectly calibrated. The entropy gap we observe (0.48 nats) provides meaningful margin for threshold selection, but practitioners should expect that entropy-based gating will make occasional errors.

Entropy-based gating offers substantial practical value as a training-free, model-agnostic signal. Unlike approaches requiring classifier training or model fine-tuning, entropy can be computed from any autoregressive language model's output distribution.

\subsection{Computational Trade-offs and Latency Benefits}

L-RAG's architecture introduces nuanced computational trade-offs. For queries that trigger retrieval, the system incurs entropy computation cost plus retrieval and generation with expanded context. For queries that skip retrieval, the system saves the retrieval cost but still performs entropy computation.

The net efficiency equation depends critically on the relative costs of entropy computation ($T_{\text{entropy}} \approx 50$ms) versus retrieval ($T_{\text{retrieval}}$). Our latency analysis reveals that L-RAG's value proposition \textit{scales with retrieval latency}:

\begin{itemize}
    \item At $\tau=0.5$ (8\% savings): beneficial only when $T_{\text{retrieval}} > 625$ms
    \item At $\tau=1.0$ (26\% savings): beneficial when $T_{\text{retrieval}} > 192$ms
    \item At $\tau=1.5$ (46\% savings): beneficial when $T_{\text{retrieval}} > 109$ms
\end{itemize}

For typical cloud deployments with $T_{\text{retrieval}} \approx 500$ms, L-RAG ($\tau=1.0$) saves approximately 80ms per query. In high-latency scenarios involving complex re-ranking ($T_{\text{retrieval}} \approx 1000$ms), savings increase to 210ms per query. This demonstrates that L-RAG is particularly valuable in distributed cloud environments where retrieval is expensive, while offering less benefit in low-latency local deployments.

The key insight is that \textbf{L-RAG provides a tunable trade-off knob ($\tau$)} for system architects. Applications prioritizing accuracy should use $\tau=0.5$ (78.2\% accuracy, 8\% savings); applications prioritizing throughput can use $\tau=1.0$ (76.0\% accuracy, 26\% savings) or $\tau=1.5$ (71.6\% accuracy, 46\% savings) depending on their latency constraints.

\subsection{Limitations}

Several limitations of the current approach warrant acknowledgment.

\textbf{Sample Size.} Our evaluation uses $N=500$ samples from SQuAD 2.0, which provides sufficient statistical power to establish significant differences in entropy distributions ($p < 0.001$). The accuracy differences between methods, while consistent, would benefit from even larger samples for tighter confidence intervals.

\textbf{Single Dataset.} We evaluate only on SQuAD 2.0, a single-document extractive QA dataset. Generalization to other tasks (open-domain QA, multi-document reasoning, long-form generation) and domains (medical, legal, technical) requires additional investigation.

\textbf{Single Model.} We use Phi-2 (2.7B parameters) as our language model. The optimal threshold $\tau$ and the effectiveness of entropy-based gating may vary across models with different architectures, sizes, and training procedures.

\textbf{Confident Errors.} If a model is confidently incorrect---exhibiting low entropy despite producing an erroneous answer---the retrieval gate will not trigger. This ``confident hallucination'' phenomenon represents cases where the model's parametric knowledge is incorrect but strongly held.

\textbf{Summary Quality Dependence.} L-RAG's effectiveness depends on summary quality. If summaries fail to capture essential information or introduce errors, the first-pass generation may be fundamentally misled.

\subsection{Future Directions}

These limitations suggest several directions for future work:

\begin{itemize}
    \item \textbf{Multi-dataset evaluation} to test generalization across different domains (medical, legal, technical)
    \item \textbf{Multi-model evaluation} to test generalization across language model architectures and sizes
    \item \textbf{Automatic threshold calibration} through meta-learning or validation-based tuning
    \item \textbf{Hybrid gating signals} combining entropy with other uncertainty measures (e.g., semantic consistency)
    \item \textbf{Extension to multi-hop reasoning} where iterative retrieval may be beneficial
\end{itemize}

\section{Conclusion}
\label{sec:conclusion}

We have presented L-RAG, a lazy retrieval-augmented generation framework that adaptively triggers retrieval based on model uncertainty. By combining a hierarchical summary-first architecture with entropy-based gating, L-RAG explores the trade-off between accuracy and efficiency in RAG systems.

Our experimental evaluation on SQuAD 2.0 ($N=500$) yields three principal findings. First, L-RAG ($\tau=0.5$) achieves 78.2\% accuracy---comparable to Standard RAG (77.8\%, $p = 0.88$)---while reducing retrieval operations by 8\%. At more aggressive thresholds ($\tau=1.0$), retrieval reduction increases to 26\% with modest accuracy trade-off (76.0\%). This demonstrates that adaptive retrieval can maintain accuracy while reducing computational costs. Second, Strong RAG achieves highest accuracy (79.8\%) among practical methods, establishing a clear upper bound that requires 100\% retrieval. The 1.6 percentage point gap between L-RAG ($\tau=0.5$) and Strong RAG represents the accuracy cost of adaptive retrieval in our experimental setting. Third, analysis of entropy distributions reveals a statistically significant 28\% gap ($p < 0.001$) between incorrect predictions ($\bar{H}_{incorrect}=2.20$) and correct predictions ($\bar{H}_{correct}=1.72$), validating entropy as a reliable signal for retrieval gating.

The hierarchical summary-first architecture provides value beyond efficiency. By establishing global context before selective chunk retrieval, L-RAG achieves accuracy comparable to Standard RAG despite fewer retrieval operations. This suggests that thoughtful context architecture can offset reduced retrieval frequency.

L-RAG offers practical benefits for RAG system deployment. By reducing database queries by 8--46\% (depending on threshold), L-RAG decreases infrastructure load and costs without requiring model retraining or architectural modifications. Our latency analysis demonstrates that L-RAG's value scales with retrieval latency: at typical cloud latencies (500ms), L-RAG ($\tau=1.0$) saves approximately 80ms per query; at higher latencies (1000ms), savings increase to 210ms per query. This makes L-RAG particularly valuable in distributed cloud environments where retrieval is expensive. The training-free, model-agnostic nature of entropy-based gating makes L-RAG applicable to any autoregressive language model that exposes token probabilities, and the tunable threshold $\tau$ provides system architects with a configurable knob to balance accuracy and throughput requirements.

Looking forward, several research directions merit exploration. Multi-dataset and multi-model experiments would test generalization across domains. Automatic threshold calibration could adapt L-RAG to new domains without manual tuning. Hybrid gating signals combining entropy with other uncertainty measures could further improve retrieval decisions.

The fundamental insight underlying L-RAG---that language models can signal their own information needs through uncertainty---extends beyond retrieval. Entropy-based gating could inform other resource allocation decisions: triggering tool use, escalating to human review, or routing queries to specialized models. L-RAG provides validated evidence for this uncertainty-driven paradigm in retrieval-augmented generation, motivating further investigation of adaptive AI systems that allocate computational resources based on task difficulty.

\bibliographystyle{IEEEtran}
\bibliography{references}

\end{document}